\documentclass{PoS}

\title{Quantum fluctuations of k-strings: A case study.}

\ShortTitle{Quantum fluctuations of k-strings: A case study.}

\author{\speaker{Pietro Giudice}, Ferdinando Gliozzi, Stefano Lottini\\
        Dipartimento di Fisica Teorica, Universit\`a di Torino and\\ 
        INFN, sezione di Torino, Italy\\

        E-mail: \email{giudice@to.infn.it},
        \email{gliozzi@to.infn.it},
        \email{lottini@to.infn.it}}

\abstract{K strings in Yang-Mills theory can be considered as bound states of
$k$ elementary confining strings carrying one unit of colour flux.
Current estimates of k-string tension $\sigma_k$ are  very sensitive
to the leading corrections due to quantum fluctuations of the string.
In this study we address this problem by comparing  Polyakov-Polyakov
correlators in the fundamental representation ($k=1$) with the corresponding
ones with $k=2$ in the confining phase of a ${\mathbb Z}_4$ gauge theory 
in three dimensions. Highly efficient simulation techniques are available 
in this case. Although the $k=1$ Polyakov-Polyakov correlator
matches nicely with the expected bosonic string effects up to the
Next-to-Leading-Order, the $k=2$ Polyakov-Polyakov correlators show large 
deviations.
This is an important source of potential systematic errors in the current
estimates of $\sigma_k$.}

\FullConference{XXIVth International Symposium on Lattice Field Theory\\
                July 23-28, 2006\\
                Tucson, Arizona, USA}

\newcommand{\bi}  {\begin{itemize}}
\newcommand{\ei}  {\end{itemize}}
\newcommand{\be}  {\begin{enumerate}}
\newcommand{\ee}  {\end{enumerate}}

\newcommand{\bc}  {\begin{center}}
\newcommand{\ec}  {\end{center}}

\newcommand{\eqa}{\begin{eqnarray}}
\newcommand{\ena}{\end{eqnarray}}

\newcommand{\bra}{\langle}
\newcommand{\ket}{\rangle}

\newcommand{\Z}  {{\mathbb Z}}
\newcommand{\eq}{\begin{equation}}
\newcommand{\en}{\end{equation}}

\begin{document}

\section{\label{sec:1} Introduction}

Significant effort has been invested recently in studies of the flux tubes 
induced by color sources in higher representations  of $SU(N)$, 
built up of 
$j$ copies of quarks in the fundamental representation. The long-distance 
properties of the flux tube  should depend  only on its $N-$ality 
$k\equiv j\,({\rm mod}\,N)$, since all representations with the same $k$ 
can be converted into each other by the emission of an appropriate number 
of soft gluons. 
 As a consequence, the heavier strings of given $N$-ality $k$
are expected to decay into the string with smallest string tension.  
The corresponding string is usually referred to as a $k$-string. If its 
tension $\sigma_k$ for any allowed $k$ satisfies the inequality 
$\sigma_{k_1+k_2}<\sigma_{k_1}+
\sigma_{k_2}$, the  $k-$string is stable against 
decay into two strings of lower $N$-ality.

Stable $k$ strings are expected to belong 
to the antisymmetric representation with $k$ quarks. 
This fact can be simply understood in terms of Casimir scaling,
 {\sl i.e.} the hypothesis that the string 
tension for a given representation is proportional to the quadratic 
Casimir operator \cite{aop,DelDebbio:1996mh}: within the set of 
all representations of $N-$ality $k$  the 
antisymmetric one corresponds to the minimum of the Casimir eigenvalues,
suggesting
\eq
\sigma_k^{(c)}=\sigma\frac{k(N-k)}{N-1}~.
\label{casi}
\en
Another competing hypothesis is the sine law:
\eq
\sigma_k^{(s)}=\sigma\frac{\sin(k\pi/N)}{\sin(\pi/N)}~,
\label{sine}
\en
which  has been derived in the large $N$ limit of ${\cal N}=2$ 
supersymmetric $SU(N)$ gauge
theory softly broken to ${\cal N}=1$ \cite {ds}, in the M theory 
description of  ${\cal N}=1$ supersymmetric $SU(N)$ gauge theory
\cite{hsz} and, more recently, in the AdS/CFT correspondence \cite{hk}. 
In some cases this formula is expected to be exact, while in others 
the calculated values of $\sigma_k$ turn out to be slightly smaller 
than $\sigma_k^{(s)}$.
 
Lattice calculations in pure $SU(N)$ gauge models for $N=6$
\cite{dprv} and $N=4,5,6,8$ \cite{lt,ltw} in $D=3+1$ point to the $k-$string 
tensions lying partway between the Casimir scaling and the sine law, however
there is no complete consensus and some dedicated studies favour the sine 
formula \cite{dprv2}. 

These calculations extract the string tensions 
$\sigma_k$ from 
Polyakov correlators at a finite temperature $T$ \emph{assuming}  
the free string prediction \cite{lsw}
\begin{equation}
\sigma_k(T)=\sigma_k-\frac{(D-2)\pi T^2}{6}+O(T^4)~,
\label{sigmat}
\end{equation} 
There is good evidence that this is a very accurate approximation to 
$k=1$ strings in all confining gauge theories irrespective of the gauge group
 once $T$ is small enough \cite{dprv,Caselle:1996ii, Gliozzi:1999wq}, but 
there is no evidence that this can be also extended for $k>1$. 
Actually we show in this work that for $k>1$ Eq.(\ref{sigmat}) is not 
adequate and needs corrections.

Our starting point was the observation that the effective string 
description, at least in the case $k=1$, 
is believed to be universal, that is to hold for all confining gauge theories.
On the other hand in order to check the validity of Eq.(\ref{sigmat}) 
it is important to do a much more accurate finite volume 
study than any currently available for $SU(N)$ $k-$strings. Hence we 
decided to study a three-dimensional $\Z_4$ gauge theory, which is  the 
simplest model supporting a $k=2$ string. Using a duality 
transformation\footnote{See the talk of S. Lottini at this conference.} 
it is possible to map this model into the symmetric Ashkin-Teller (AT) 
model, where very high precision can be achieved on large lattices, through
a non-local cluster algorithm. The outcome of this analysis is 
uncontroversial: though at low temperature the tension of the fundamental 
string fits nicely with Eq.(\ref{sigmat}) even at Next-to-Leading-Order
(NLO), we find a clear mismatch for the 2-string 
in the same temperature range, suggesting the need of corrections to 
Eq.(\ref{sigmat}). 
 
Theoretically, there is something one can say about  the origin of these corrections. $k-$strings can be viewed as  bound states of $k$ fundamental strings. 
Accordingly, we expect that besides the mechanical vibration modes whose 
quantum contributions yield exactly Eq.(\ref{sigmat}), there should be 
breathing modes related to the internal degrees of freedom of the $k$ 
constituent strings.       

\section{\label{sec1b} The effective string model}
The infrared description of any confining gauge theory is well described by 
an effective string model.
A particularly simple string action is the Nambu-Goto one, where  
the correlation function of two Polyakov 
loops at a temperature $T=1/L$ and at a distance $R$ can be calculated at the
NLO \cite{difi83,cahapa05}, yielding
\eq
\bra P(0)\,P^\dagger(R)\ket\propto
 \frac{{\rm e}^{-cL-\sigma \,RL+\frac{(D-2)\pi^2 L\,E(\tau)}
{1152\sigma\, R^3}+O(1/R^5)}}{\eta(\tau)^{D-2}}
~;~\tau\equiv\frac{iL}{2R}~;~E=2E_4-E^2_2~,
\label{nambu-goto}
\en
where $\eta$ is the Dedekind eta function and $E_n(\tau)$ are the Eisenstein
functions (see e.g. \cite{cahapa05} for detailed definitions).
There is strong evidence that at this order this formula is universal, 
i.e. it holds in whatever confining gauge theory 
\cite{luwe04,Drummond:2004yp,dama06}.

\section{\label{sec:2} Algorithm}
We work directly in the dual form of the $3d \, \Z_4$ gauge model, i.e.
a symmetric Ashkin-Teller (AT) model. 
It  is described in terms of two coupled, ferromagnetic, Ising systems through 
the two-parameter action
\eq
S=-\sum_{\langle xy \rangle} \beta \ (\sigma_x \sigma_y + \tau_x \tau_y) +
\alpha \ \sigma_x \sigma_y \tau_x \tau_y,
\en
where $\sigma_x$ and $\tau_x$ are the Ising variables 
($\sigma_x,\tau_x = \pm 1$).
The global $\Z_4$ symmetry of the action is generated by the 
transformation  $\sigma \rightarrow -\tau, \ \tau \rightarrow \sigma$.
An independent  ${\mathbb Z}_2$ symmetry is generated by the transformation 
$\sigma \leftrightarrow \tau$, which is related to the charge conjugation 
of the corresponding dual model. 

The great advantage of studying the $3d$ AT model instead of the original 
$3d \, \Z_4$ gauge theory is that a non-local 
cluster updating algorithm~\cite{Swendsen:1987ce} can be used.
Of course, in the AT model we have two spin variables, so it is necessary 
to extend the original method.
The idea is the following: we take the two site variables 
$\sigma$ and $\tau$ as if they belonged to two distinct lattices $R_\sigma$ and
$R_\tau$; we freeze the lattice $R_\sigma$ and  apply the 
Swendsen-Wang algorithm to the variables $\tau$ so obtaining a new lattice
$R^\prime_\tau$. At this point we freeze the lattice $R^\prime_\tau$ and 
we update the variables $\sigma$, and so on.

It is possible to show that in this statistical system the expectation 
value of the Wilson loop of the gauge model
$\langle W_\gamma \rangle_{gauge}$ is given by:
$$
\langle W_\gamma \rangle_{gauge}= \frac{Z^*_{AT}}{Z_{AT}},
$$
where $Z_{AT}$ is the partition function of the AT model and $Z^*_{AT}$
is that modified by a suitable twist of the couplings.
In particular, in order to determine the $\langle W_\gamma 
\rangle_{gauge}$ in 
the fundamental representation $(k=1)$ it suffices to flip the 
couplings of  $\sigma$ {\sl or} $\tau$ just in the links orthogonal to
an arbitrary surface $\Sigma$ encircled by $\gamma=\partial\,\Sigma$.
Similarly, flipping the signs of {\sl both} couplings of
$\sigma$  and $\tau$ in the same surface,
we get the Wilson loop in the $k=2$ representation.

Actually, since we have a model written in terms of Ising variables 
we can use a very powerful method to estimate the expectation value of the
Wilson loop based only on the topological linking  properties of 
the Fortuin-Kasteleyn (FK) clusters~\cite{Gliozzi:1996fy}. 
For each FK configuration one looks 
for paths in the clusters and then applies the following rule:
\bi
\item $W_\gamma=1$ if there is no path linked with the loop $\gamma$ or 
if the winding number modulo 2 is zero,
\item  $W_\gamma=0$ otherwise.
\ei
Note that using such a definition, the value of $W_\gamma$ does not change if
dangling ends or bridges between closed path are added or 
removed~\cite{Gliozzi:2005ny}. 

In the AT model we apply the above rules separately for the two variables 
$\sigma$ and $\tau$, therefore we have, for each configuration, 
a value $W^\sigma$ and a value 
$W^\tau$, each of them corresponds to the value of the Wilson loop 
in the fundamental representation; the product $W^\sigma W^\tau$ corresponds
to the value of Wilson loop in the $k=2$ representation.
 
The same ideas do apply to measure the Polyakov-Polyakov correlator
$\langle P(0) P^\dagger(R) \rangle$.


\section{\label{sec:3} Numerical results}

In this paper we discuss data obtained by simulating the AT model in 
one point of the parameter space $(\alpha=0.05,\beta=0.207)$, where our 
best estimate for string tensions are
$\sigma_1=0.02084(5)$ and $\sigma_2=0.0323(5)$.
In this case, the tension ratio is
$\sigma_2/\sigma_1 \simeq 1.55$, while the values predicted by the sine 
or Casimir scaling for $N=4$ and $k=2$ are $\sqrt{2} \simeq 1.41$ 
and  $4/3$, respectively.

We performed $10^6$ measurements for both $\sigma_1$ and $\sigma_2$ on 
a lattice $64^2 \times L$. The value of $L$ has been determined 
in such a way  that $T \simeq T_c/2$ using the relation 
$T_c/\sqrt{\sigma}\simeq 1.1$.

In order to check whether the functional relation (\ref{nambu-goto}) 
 is an accurate description of the Polyakov-Polyakov correlator  
two conditions are needed:
the fitting parameters (in particular $\sigma_k$) {\it should be stable}
at large $R$ and
the estimated value of  $\sigma_k$ {\it should not depend on $T$}
since Eq.(\ref{nambu-goto}) should account for the $T$ dependence.

\begin{figure}[ht]
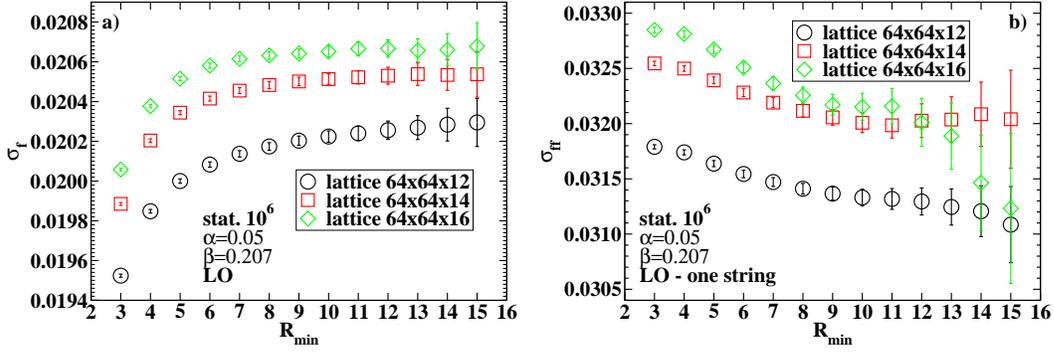

\begin{minipage}{70mm}
\bc
\includegraphics[angle=0,width=6.8cm]{./figure/graf_sigma_vd_index-LO-s1.eps}
\ec
\end{minipage}
%
%
\begin{minipage}{70mm}
\bc
\includegraphics[angle=0,width=6.8cm]{./figure/graf_sigma_vd_index-LO-s2.eps}
\ec
\end{minipage}
\vspace{-2mm}
\caption{$\sigma_f$ and $\sigma_{ff}$ vs $R_{min}$ for three different $T$ 
at LO}
\label{fig2}
\end{figure}

In Fig.~\ref{fig2}a it is possible to see the $\sigma_f$ value\footnote{In 
this Section $\sigma_f \equiv  \sigma_1$ and $\sigma_{ff} \equiv  
\sigma_{2}$.} fitted
by the Leading-Order (LO) approximation (i.e. neglecting the $E$ term in 
(\ref{nambu-goto})) in a range $[R_{min},18]$ 
(where $R_{min}$ is the value that appears on the x-axes);
albeit for large values of $R_{min}$ it seems to appear a plateau, 
there are three
different values of $\sigma_f$, therefore the  functional relation (FR) used
is not sufficiently accurate to determine it.

Fig.~\ref{fig2}b is similar to Fig.~\ref{fig2}a, but in this case we have 
used the same FR to determine the value of  $\sigma_{ff}$; also in this 
case we can apply the previous considerations and the FR is not correct.

\begin{figure}[ht]
\bc
\includegraphics[angle=0,width=8cm]{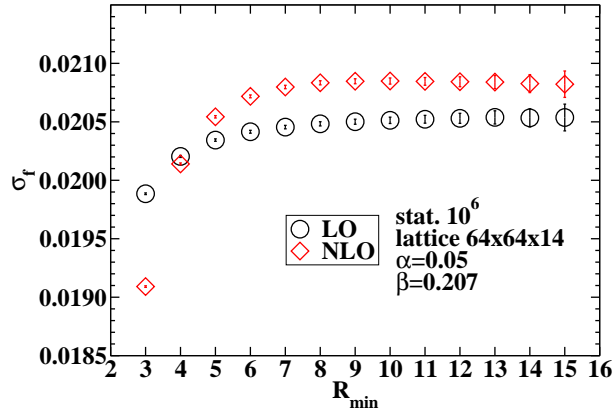}
\vspace{-4mm}
\caption{$\sigma_f$ vs $R_{min}$ for LO and NLO}
\label{fig3}
\ec
\end{figure}

Then, in Fig.~\ref{fig3}, we fitted the value of $\sigma_f$ to two 
different FRs,
one obtained by a LO approximation and the other one by NLO;
the value of $\sigma_f$ increases significantly and the plateau appears for 
a smaller value of $R_{min}$.

\begin{figure}[ht]
\bc
\includegraphics[angle=0,width=8cm]{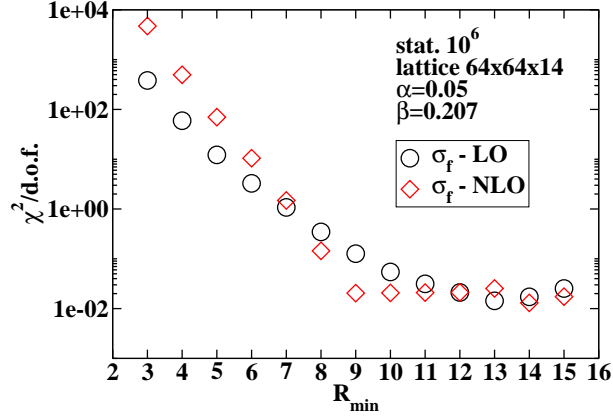}
\vspace{-4mm}
\caption{$\chi^2_{\sigma_f}/\mbox{d.o.f}$ vs $R_{min}$ for LO and NLO}
\label{fig4}
\ec
\end{figure}

In order to shed some light on the difficulty in determining a string 
tension  that is 
free of systematic errors, in Fig.~\ref{fig4} we plot the values of 
$\chi^2/d.o.f.$ for fits which appear in Fig.~\ref{fig3};
note that for $R_{min} \geq 11$ there is no difference in $\chi^2/d.o.f.$ 
between LO and NLO, even if
the values of $\sigma_f$ in the same range (see Fig.~\ref{fig3}) are different.
It is interesting to note that in NLO case also the value of  $\chi^2/d.o.f.$
shows a plateau for $R_{min} \geq 9$, therefore this FR better describes
what happens at small values of $R$.

\begin{figure}[ht]
\bc
\includegraphics[angle=0,width=8cm]{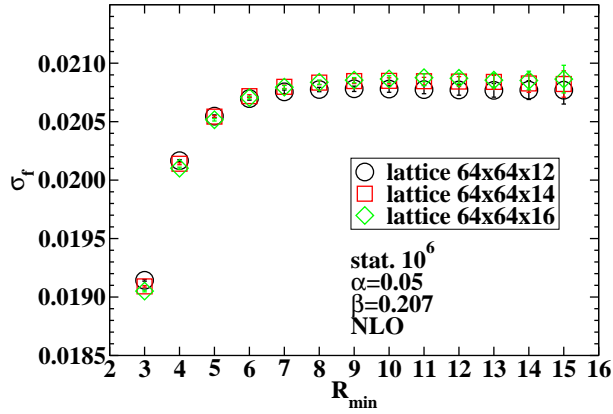}
\vspace{-4mm}
\caption{$\sigma_f$ vs $R_{min}$ for three different $T$ at NLO}
\label{fig5}
\ec
\end{figure}

Fig.~\ref{fig5} is similar to Fig.~\ref{fig2}a but now we use FRs 
obtained with a NLO approximation; now the FR is actually correct: 
the value of $\sigma_f$ 
is stable also for small values of $R_{min}$ and there is only one value
of it for three different values of $L$ (i.e. different values of $T$).

\begin{figure}[ht]
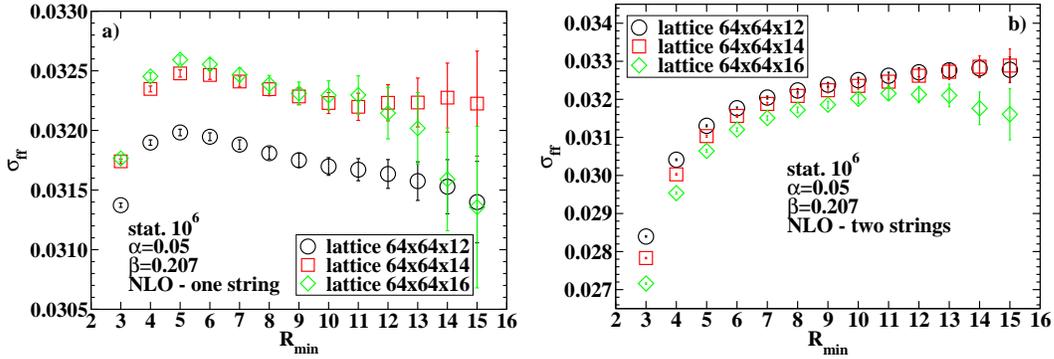

\begin{minipage}{70mm}
\bc
\includegraphics[angle=0,width=6.8cm]{./figure/graf_sigma_vd_index-NLO-s2.eps}
\ec
\end{minipage}
%
%
\begin{minipage}{70mm}
\bc
\includegraphics[angle=0,width=6.8cm]{./figure/graf_sigma_vd_index-NLO-s2-twostrings.eps}
\ec
\end{minipage}
\vspace{-2mm}
\caption{$\sigma_{ff}$ vs $R_{min}$ for three different $T$ at NLO: a) one string; b) two strings}
\label{fig7}
\end{figure}

Finally, in Fig.~\ref{fig7}a and Fig.~\ref{fig7}b we interpolate the value of 
$\sigma_{ff}$ with NLO 
approximation with two different hypotheses: in Fig.~\ref{fig7}a the two strings
are stuck together and fluctuate as a single string (we put D=3 in 
(\ref{nambu-goto})); in Fig.~\ref{fig7}b the
two strings are assumed to fluctuate independently (this is equivalent to 
putting D=4 in (\ref{nambu-goto})). It is clear that in this case 
the FR  does not describe  accurately $\sigma_{ff}$:
in both cases a plateau does not appear for $\sigma_{ff}$ and there are
different values for different values of $T$.
This indicates that the current estimates of $\sigma_{k}$ are affected 
by systematic errors.

\section{Conclusions}

The functional relations which are very accurate to fit
data related to fundamental string tension {\it are not adequate} to
describe 2-string tension.

An accurate estimate of the $k$-string tension is rather problematic:
from a numerical point of view our analysis shows that a blind application 
of the usual 
formulas for the Polyakov-Polyakov correlators introduces strong 
systematic errors.
From a theoretical point of view it is necessary to face the problem
of determining the correct functional relation of Polyakov-Polyakov correlator
for higher representations, which is presently unknown.

\end{document}